\documentclass[a4paper,10pt,twoside]{cpc-hepnp}
\usepackage{CJK,upgreek,fancyhdr}
\usepackage{multicol}
\usepackage{graphicx}
\usepackage{booktabs}
\usepackage{amssymb,bm,mathrsfs,bbm,amscd}
\usepackage[tbtags]{amsmath}
\usepackage{lastpage}

\begin{document}
\begin{CJK*}{GB}{gbsn}

\fancyhead[c]{\small Chinese Physics C~~~Vol. xx, No. x (201x) xxxxxx}
\fancyfoot[C]{\small 010201-\thepage}

\footnotetext[0]{Received 31 June 2015}

\title{Effects of entrance channel on fusion probability in hot fusion reactions\thanks{Supported by the Young Teacher Start Programme of Sun Yat-Sun University (45000-18821200), The Natural Science Foundation of Guangdong Province under Grant Nos. 2016A030310208, the Fundamental Research Funds for the Central Universities under Grant No. 15lgpy30, and the National Natural Science Foundation of China under Grant Nos. 11405278.}}

\author{%
    Long Zhu (×£Áú)$^{1;1)}$\email{zhulong@mail.sysu.edu.cn}%
\quad Jun Sun (ËÕ¾ü)$^{1}$
\quad Ching-Yuan Huang (»ÆÇìÔª)$^{1}$
\quad Feng-Shou Zhang (ÕÅ·áÊÕ)$^{2,3;2)}$\email{fszhang@bnu.edu.cn}%
}
\maketitle

\address{%
$^1$ Sino-French Institute of Nuclear Engineering and Technology, Sun Yat-sen University, Zhuhai 519082, China\\
$^2$ The Key Laboratory of Beam Technology and Material Modification of Ministry of Education, College of Nuclear Science and Technology, Beijing Normal University, Beijing 100875, China\\
$^3$ Beijing Radiation Center, Beijing 100875, China
}

\begin{abstract}
Within the framework of the dinuclear system (DNS) model, the fusion reactions leading to the compound nuclei $^{274}$Hs$^{*}$ and $^{286}$Cn$^{*}$ are investigated. The fusion probability as a function of DNS excitation energy is studied. The calculated results are in good agreement with the available experimental data. The obtained results show that the fusion probabilities are obviously enhanced for the reactions located at high place in potential energy surface, although these reactions may have small values of mass asymmetry. It is found that the enhancement is due to the large potential energy of the initial DNS.
\end{abstract}



\footnotetext[0]{\hspace*{-3mm}\raisebox{0.3ex}{$\scriptstyle\copyright$}2016
Chinese Physical Society and the Institute of High Energy Physics
of the Chinese Academy of Sciences and the Institute
of Modern Physics of the Chinese Academy of Sciences and IOP Publishing Ltd}%

\begin{multicols}{2}

\section{Introduction}

Production of superheavy nuclei (SHN) is one of most important research areas in nuclear physics. Heavy ion fusion reactions have been extensively investigated in synthesis of SHN experimentally and theoretically.
The SHN with $Z=102-118$ have been synthesized in the $^{208}$Pb- and $^{209}$Bi-based cold fusion reactions \cite{Hofmann01,Kosuke01} and $^{48}$Ca-induced hot fusion reactions \cite{Yu01,Yu02,Yu04}.
Recently, an experiment has been performed to synthesize the superheavy element (SHE) $Z=120$ using the reaction $^{244}$Pu($^{58}$Fe,xn)$^{302-x}$120 at FLNR (Dubna) \cite{Yu03}. Unfortunately, no correlated decay chains were observed. Cap $\emph{et al}$. predicted that it is impossible to synthesize SHEs with $Z=122$ and 124 in the fusion of symmetric systems \cite{Cap01}. Hence, the choice of more mass asymmetric reactions would be preferable. Choosing favorable reactions is very important for sythesizing SHEs with $Z>118$. Therefore, it is worth investigating the mechanism of heavy ion fusion reactions theoretically.

The evaporation residual (ER) cross sections of fusion reactions strongly depend on the projectile-target combination. A lot of works have been done to investigate the entrance channel effects on production of SHN \cite{Nasirov01,Fazio01,Kim01,Hong01,Nasirov02,Zagrebaev01,Shen01,Wang02}, including the reaction Q value and orientation effects on the ER cross sections \cite{Liu02,Nishio02,Zhu02,Umar01}. However, the mechanism of entrance channel effects on productions of SHN, especially entrance channel effects of the fusion probability ($P_{\textrm{CN}}$), is still not clear. Fusion probability is the least well known quantity that determines the ER cross sections. It is known that the fusion probability of massive nuclei strongly depends on the mass (charge) asymmetry $\eta=(A_{1}-A_{2})/(A_{1}+A_{2})$ ($\eta_{Z}=(Z_{1}-Z_{2})/(Z_{1}+Z_{2})$), where $A_{1}$, $A_{2}$ and $Z_{1}$, $Z_{2}$ are the mass and charge numbers of two fragments. Usually, $P_{\textrm{CN}}$ is larger for a mass asymmetric reaction than that for the symmetric one, because of the lower inner fusion barrier $B_{\textrm{fus}}$ and the higher quasifission barrier $B_{\textrm{qf}}$ \cite{Adamian04}. But an opposite behavior can be seen for the reactions $^{124}$Sn + $^{92}$Zr and $^{86}$Kr + $^{130}$Xe in Ref. \cite{Fazio01}. Therefore, it is desirable to get more detailed studies about the effects of entrance channel on $P_{\textrm{CN}}$ in synthesis of SHN.

Comparison of $P_{\textrm{CN}}$ in different mass asymmetry reactions leading to the same compound nucleus (CN) allows us to analyze the entrance channel effects on the fusion-fission reaction mechanism in collisions of massive nuclei. The DNS model has been successfully used in investigating the synthesis mechanism of SHN \cite{Feng01,Feng02,Huang01,Adamian01,Adamian02,Feng03,Adamian04,Adamian05,Adamian06,Bao01,Bao02,Wang01}. In this work, we focus on the entrance channel effects of $P_{\textrm{CN}}$ in synthesis of SHN within the framework of DNS model. The ER cross sections and $P_{\textrm{CN}}$ for the reactions $^{26}$Mg + $^{248}$Cm, $^{36}$S + $^{238}$U, and $^{48}$Ca + $^{226}$Ra are investigated. The fusion probabilities leading to the compound nucleus $^{286}$Cn in various reactions $^{48}$Ca + $^{238}$U, $^{52}$Ca + $^{234}$U, $^{53}$Ca + $^{233}$U, and $^{55}$Sc + $^{231}$Pa are investigated.

The article is organized as follows. In Sec. 2, we describe the DNS model. The results and discussion of $P_{\textrm{CN}}$ are presented in Sec. 3. Finally, we summarize the main results in Sec. 4.

\section{Description of the model}

The ER cross sections of SHN can be written as a sum over all partial waves $J$,
\begin{gather}
\begin{split}
\sigma_{\textrm{ER}}(E_{\textrm{c.m.}})=&\frac{\pi\hbar^{2}}{2\mu E_{\textrm{c.m.}}}\sum_{J}(2J+1)T(E_{\textrm{c.m.}},J)\\
&\times P_{\textrm{CN}}(E_{\textrm{c.m.}},J)W_{\textrm{sur}}(E_{\textrm{c.m.}},J).
\end{split}
\end{gather}
Here, $T$, $P_{\textrm{CN}}$, and $W_{\textrm{sur}}$ are transmission probability, fusion probability, and survival probability, respectively. $E_{\textrm{c.m.}}$ and $J$ are the incident energy and incident angular momentum in the center of mass frame, respectively. In this work, the transmission probability $T$ in the capture process is calculated
by empirical coupled-channel model \cite{Feng02}, in which the Woods-Saxon potential is used for nuclear potential, and its parameters $V_{0}=80$ MeV and $a=0.68$ fm, are adopted.

The fusion probability is investigated within DNS model. In this work, the diffusion process is treated along proton and neutron degrees of freedom. The probability distribution function $P(Z_{1},N_{1},E_{1},t)$ for fragments 1 with proton number $Z_{1}$, neutron number $N_{1}$ and excitation energy $E_{1}$ at time $t$ can be obtained by solving the following master equations \cite{Feng03},
\begin{flalign}
\begin{split}\label{master}
&\frac{dP(Z_{1},N_{1},E_{1},t)}{dt}\\
&=\sum_{Z_{1}^{'}}W_{Z_{1},N_{1};Z_{1}^{'},N_{1}}(t)[d_{Z_{1},N_{1}}P(Z_{1}^{'},N_{1},E_{1}^{'},t)\\
&-d_{Z_{1}^{'},N_{1}}P(Z_{1},N_{1},E_{1},t)]\\
&+\sum_{N_{1}^{'}}W_{Z_{1},N_{1};Z_{1},N_{1}^{'}}(t)[d_{Z_{1},N_{1}}P(Z_{1},N_{1}^{'},E_{1}^{'},t)\\
&-d_{Z_{1},N_{1}^{'}}P(Z_{1},N_{1},E_{1},t)]\\
&-\Lambda_{\textrm{qf}}(\Theta(t))P(Z_{1},N_{1},E_{1},t).
\end{split}
\end{flalign}
Here $W_{Z_{1},N_{1};Z_{1}^{'},N_{1}}$ ($W_{Z_{1},N_{1};Z_{1},N_{1}^{'}}$) denotes the mean transition probability from the channel ($Z_{1}$, $N_{1}$, $E_{1}$) to ($Z_{1}^{'}$, $N_{1}$, $E_{1}^{'}$) [or ($Z_{1}$, $N_{1}$, $E_{1}$) to ($Z_{1}$, $N_{1}^{'}$, $E_{1}^{'}$)], and $d_{Z_{1},N_{1}}$ is
the microscopic dimension corresponding to the macroscopic state ($Z_{1}$, $N_{1}$, $E_{1}$). In the DNS model, we consider the process of only one nucleon transfer. The sum is taken over all possible proton and neutron numbers that fragment 1 may take.
$\Lambda_{\textrm{qf}}$ is the quasifission rate, which describes the evolution of DNS system along relative distance R. The detailed description of $\Lambda_{\textrm{qf}}$ can be seen in Ref. \cite{Zhu02}.

The local excitation energy $E_{1}$ in Eq. (\ref{master}) is related to the excitation energy of the composite system at the initial state $E^{*}_{\textrm{DNS}}$ and the driving potentials of the injection point of the DNS. $E_{1}$ is given by
\begin{flalign}
\begin{split}\label{local}
E_{1}&=E^{*}_{\textrm{DNS}}-U(Z_{1}, N_{1}, Z_{2}, N_{2}, J)\\
&+U(Z_{\textrm{p}}, N_{\textrm{p}}, Z_{\textrm{t}}, N_{\textrm{t}}, J)-\frac{(J-M)^{2}}{2\zeta_{\textrm{rel}}}-\frac{M^{2}}{2\zeta_{\textrm{int}}},
\end{split}
\end{flalign}
where $U(Z_{1}, N_{1}, Z_{2}, N_{2})$ and $U(Z_{\textrm{p}}, N_{\textrm{p}}, Z_{\textrm{t}}, N_{\textrm{t}})$ are the potential energies of the configuration ($Z_{1}$, $N_{1}$; $Z_{2}$, $N_{2}$) and the DNS at the injection point. $M$ denotes the intrinsic angular momentum derived from the dissipation of the relative angular momentum, and $\zeta_{\textrm{int}}$ is the corresponding moment of inertia. $J$ denotes the initial angular momentum. $\zeta_{\textrm{rel}}$ is the relative moment of inertia of the DNS, which is given by $\zeta_{\textrm{rel}}=\mu R^{2}_{\textrm{m}}$.
The local excitation energy means the excitation energy of the DNS for a specific configuration ($Z_{1}$, $N_{1}$; $Z_{2}$, $N_{2}$). $E^{*}_{\textrm{DNS}}$ is converted from the relative kinetic energy loss, which is
related to the incident energy and the minimum of the well bottom of the nucleus-nucleus
potential in entrance channel $V(R_{\textrm{m}})$ \cite{Wolschin01}. Here, $R_{\textrm{m}}$ is the distance of two colliding nuclei located at the bottom of the potential pocket.
$E^{*}_{\textrm{DNS}}$ can be written as,
\begin{flalign}
\begin{split}
E^{*}_{\textrm{DNS}}=E_{\textrm{c.m.}}-V(R_{\textrm{m}})-\frac{(J\hbar)^{2}}{2\zeta_{\textrm{rel}}}.
\end{split}
\end{flalign}

The potential energy surface (PES) of the DNS is given by
\begin{flalign}
\begin{split}
U&(Z_{1}, N_{1}, R_{\textrm{m}}, J)=B(Z_{1}, N_{1})+B(Z_{2}, N_{2})\\
&-B(Z, N)-V^{\textrm{CN}}_{\textrm{rot}}(J)+V_{\textrm{CN}}(Z_{1}, N_{1}, R_{\textrm{m}}, J).
\end{split}
\end{flalign}
$B(Z_{i}, N_{i})$ ($i=1$, 2) and $B(Z, N)$ are the ground state binding energies of the fragment $i$ and compound nucleus ($Z=Z_{1}+Z_{2}$; $N=N_{1}+N_{2}$), respectively. $V_{\textrm{CN}}$ is the interaction potential of two fragments. The details of $V_{\textrm{CN}}$ are given in Ref. \cite{Feng01}.
The $Q_{\textrm{gg}}$ value of forming a configuration ($Z_{1}$, $N_{1}$; $Z_{2}$, $N_{2}$) is defined as
\begin{flalign}
\begin{split}
Q_{\textrm{gg}}=B_{\textrm{DNS}}(Z_{\textrm{p}}, N_{\textrm{p}}; Z_{\textrm{t}}, N_{\textrm{t}})-B_{\textrm{DNS}}(Z_{1}, N_{1}; Z_{2}, N_{2}).
\end{split}
\end{flalign}
Here, $B_{\textrm{DNS}}$ is the binding energy of DNS system, which can be written as $B_{\textrm{DNS}}(Z_{1}, N_{1}; Z_{2}, N_{2})=B(Z_{1}, N_{1})+B(Z_{2}, N_{2})$.

\end{multicols}

\begin{center}
\tabcaption{ \label{tab:table1}Physical quantities (the quasifission barrier $B_{\textrm{qf}}$, the inner fusion barrier $B_{\textrm{fus}}$, V($R_{\textrm{m}}$), and Coulomb barrier V$_{B}$) for the indicated reactions. Angular momentum $J=0$.}
\footnotesize
\begin{tabular*}{160mm}{c@{\extracolsep{\fill}}cccccc}

   \toprule    Reactions & Compound    & $B_{\textrm{qf}}$ & $B_{\textrm{fus}}$ & V($R_{\textrm{m}}$) & V$_{B}$\\

                 $ $    & nuclei & (MeV) & (MeV) & (MeV) & (MeV) \\
      \hline
                 $^{26}$Mg + $^{248}$Cm & $^{274}$Hs$^{*}$ &  10.72 &  & 111.80 & 122.52 \\

                 $^{36}$S + $^{238}$U   & $^{274}$Hs$^{*}$ &  8.96 & 2.71 & 144.12 & 153.08 \\

                 $^{48}$Ca + $^{226}$Ra  & $^{274}$Hs$^{*}$ & 6.65 & 6.82  & 169.80  & 176.45 \\

                 $^{48}$Ca + $^{238}$U  & $^{286}$Cn$^{*}$ &  6.25 & 10.29    & 177.13 & 183.38  \\

                 $^{52}$Ca + $^{234}$U  & $^{286}$Cn$^{*}$ &  6.60 &  9.08     &  175.24  & 181.84 \\

                 $^{53}$Ca + $^{233}$U  & $^{286}$Cn$^{*}$ &  6.70 &   7.86    &  174.83  & 181.53\\

                  $^{55}$Sc + $^{231}$Pa  & $^{286}$Cn$^{*}$ & 5.86 &  10.47    &  182.39 & 188.25 \\

      \bottomrule
\end{tabular*}%
\end{center}

\begin{multicols}{2}

The fusion probability is expressed as follows:
\begin{flalign}
\begin{split}
P_{\textrm{CN}}(E_{\textrm{c.m.}},J)=\sum_{A_{1}=1}^{A_{\textrm{BG}}}P(A_{1},E_{\textrm{c.m.}},J),
\end{split}
\end{flalign}
where $A_{\textrm{BG}}$ (=$N_{\textrm{BG}}+Z_{\textrm{BG}}$) is the mass number of light fragment at Businaro-Gallone (B.G.) point.

The survival probability of emitting $x$ neutrons can be written as
\begin{flalign}
\begin{split}
W_{\textrm{sur}}(E^{*}_{\textrm{CN}}, x, J)=&P(E^{*}_{\textrm{CN}}, x, J)\times\\
&\prod_{i}^{x}[\frac{\Gamma_{n}(E_{i}^{*}, J)}{\Gamma_{n}(E^{*}_{i}, J)+\Gamma_{f}(E^{*}_{i},J)}].
\end{split}
\end{flalign}
Here $E^{*}_{\textrm{CN}}$, $J$ are the excitation energy and the spin of the compound nucleus, respectively. $E^{*}_{i}(=E^{*}_{i-1}-B_{i-1}^{n}-2T_{i-1}$) is the excitation energy before evaporation of the $i$th neutron.
The detailed description of the width of the $i$th neutron emission and fission can be found in Ref. \cite{Feng02}.
The realization probability $P(E^{*}_{\textrm{CN}}, x, J)$ is calculated as in Ref. \cite{Hong01}. In this work, the fission barrier before evaporating the $i$th neutron is obtained by
\begin{flalign}
\begin{split}
B^{f}_{i}(E_{i}^{*})=B^{f}_{\textrm{M}}(E^{*}_{i}=0)\textrm{exp}(-E^{*}_{i}/E_{d}).
\end{split}
\end{flalign}
The microscopic shell correction energy $B^{f}_{\textrm{M}}(E^{*}_{i}=0)$ is taken from Ref. \cite{Moller01}. The damping factor $E_{d}=20$ MeV is taken.

\section{Results and discussion}

\subsection{Reactions to the compound nucleus $^{274}$Hs$^{*}$}

The DNS model is used to investigate the fusion reactions $^{26}$Mg + $^{248}$Cm, $^{36}$S + $^{238}$U, and $^{48}$Ca + $^{226}$Ra leading to the same compound nucleus $^{274}$Hs$^{*}$.

\begin{center}
\includegraphics[width=6cm]{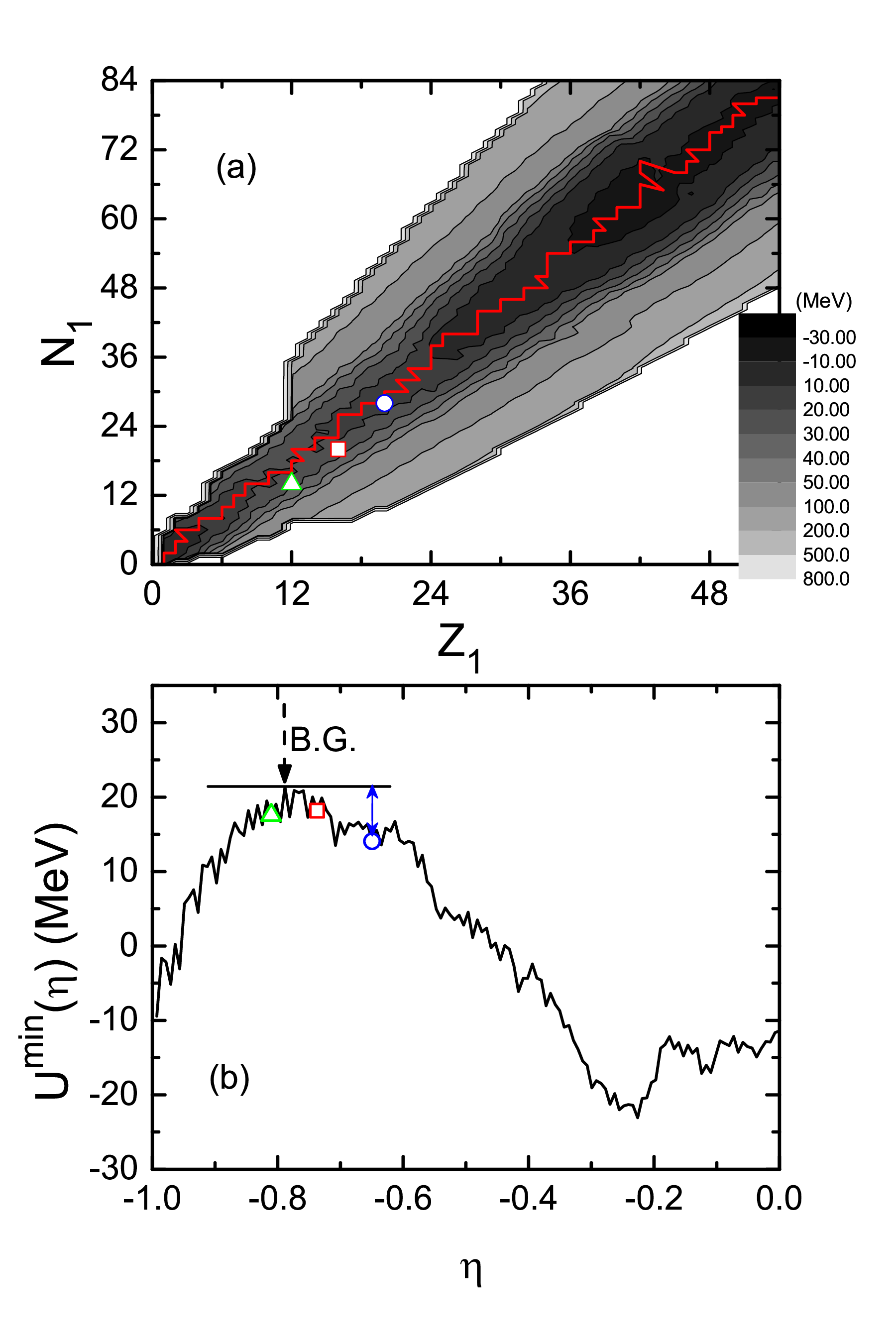}
\figcaption{\label{driver274Hs} (Color online) (a) The PES as functions of $Z_{1}$ and $N_{1}$ of the fragment 1 for the DNS configurations leading to the compound nucleus $^{274}$Hs$^{*}$. The red line indicates the minimum trajectory in the PES. Triangle, square, and circle represent the injection points for the reactions $^{26}$Mg + $^{248}$Cm, $^{36}$S + $^{238}$U, and $^{48}$Ca + $^{226}$Ra, respectively. (b) The driving potential, $U^{\textrm{min}}(\eta)$, which is the line connecting the minimum trajectory as a function of mass asymmetry $\eta$. The dashed arrow denotes the position of the B.G. point. The vertical solid arrow indicates the inner fusion barrier for the reaction $^{48}$Ca + $^{226}$Ra.}
\end{center}

Figure \ref{driver274Hs}(a) shows the PES as functions of proton and neutron number of the fragment 1 for the DNS configurations in formation of the compound nucleus $^{274}$Hs$^{*}$. The red line indicates the
minimum trajectory in the PES. We show injection points in PES for the reactions $^{26}$Mg + $^{248}$Cm, $^{36}$S + $^{238}$U, and $^{48}$Ca + $^{226}$Ra, which are all close to the red line. The minimum trajectory in PES ($U^{\textrm{min}}$) can be presented as a function of the mass asymmetry $\eta$, as shown in Fig. \ref{driver274Hs}(b). The dashed arrow denotes the position of the B.G. point. The vertical solid arrow indicate the inner fusion barrier $B_{\textrm{fus}}$ for the reaction $^{48}$Ca + $^{226}$Ra. Usually, the inner fusion barrier $B_{\textrm{fus}}$ and quasifission barrier $B_{\textrm{qf}}$ play a main role in the competition between quasifission and complete fusion \cite{Fazio01,Adamian04}.

\begin{center}
\includegraphics[width=6cm]{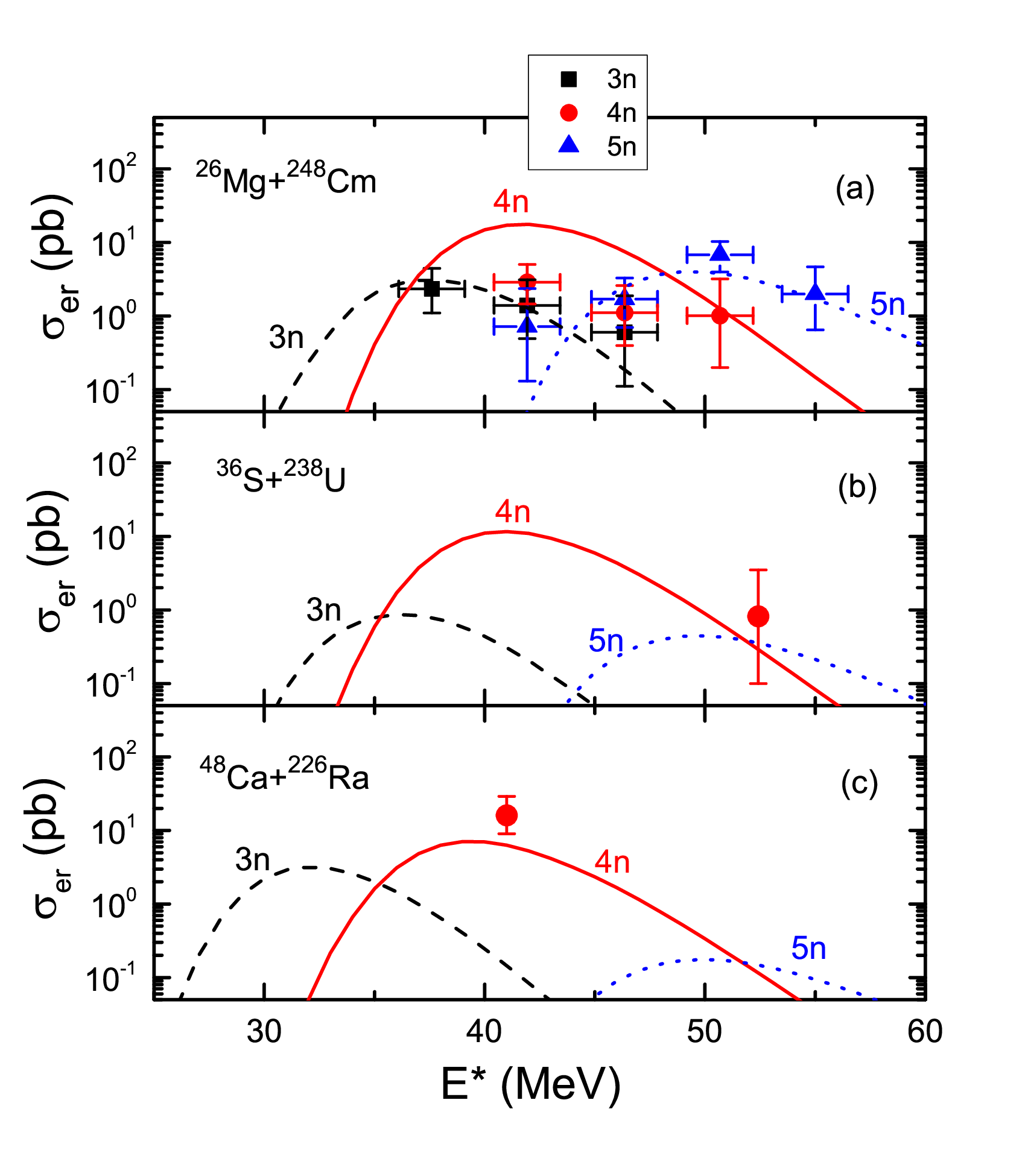}
\figcaption{\label{cross sections} (Color online) Comparison of the calculated ER cross sections with the available experimental data \cite{Dvorak01,Graeger01,Yu02} for the reactions $^{26}$Mg + $^{248}$Cm (a), $^{36}$S + $^{238}$U (b), and $^{48}$Ca + $^{226}$Ra (c). The experimental data of 3n, 4n, and 5n channels are denoted by squares, circles, and triangles, respectively.}
\end{center}

\begin{center}
\includegraphics[width=6cm]{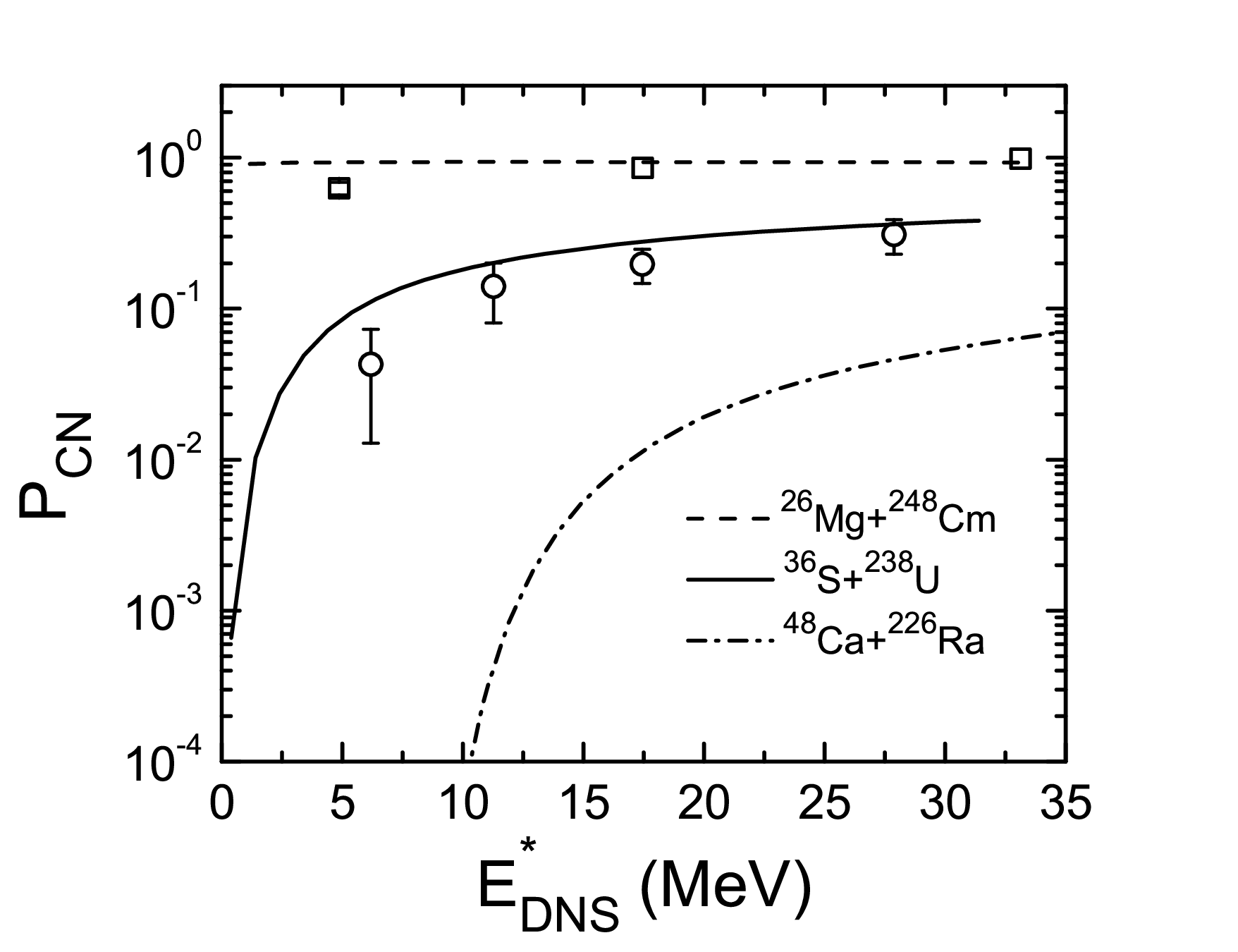}
\figcaption{\label{P-Hs} Calculated fusion probabilities as a function of $E_{\textrm{DNS}}^{*}$ ($J=0$) for the reactions $^{26}$Mg + $^{248}$Cm, $^{36}$S + $^{238}$U, and $^{48}$Ca + $^{226}$Ra. Experimental values of $P_{\textrm{CN}}$ obtained in Ref. \cite{Itkis01} for the reactions $^{26}$Mg + $^{248}$Cm and $^{36}$S + $^{238}$U are denoted by squares and circles, respectively.}
\end{center}

Figure \ref{cross sections} shows the ER cross sections for the reactions $^{26}$Mg + $^{248}$Cm, $^{36}$S + $^{238}$U, and $^{48}$Ca + $^{226}$Ra. The calculated ER cross sections are in agreement with the experimental data within a factor of 1-3. For the reaction $^{36}$S + $^{238}$U, we predict that the maximal ER cross section of 4n channel is 11.6 pb, and the corresponding excitation energy equals $E^{*}=41$ MeV.

Figure \ref{P-Hs} presents the dependence of the fusion probabilities on excitation energy of DNS for the reactions $^{26}$Mg + $^{248}$Cm, $^{36}$S + $^{238}$U, and $^{48}$Ca + $^{226}$Ra in formation of same compound nucleus $^{274}$Hs$^{*}$. The $P_{\textrm{CN}}$ is calculated with $J=0$ \cite{Zagrebaev01}. The available experimental data for the reactions $^{26}$Mg + $^{248}$Cm and $^{36}$S + $^{238}$U are also shown. It can be seen from the experimental data that the fusion probabilities for the reaction $^{26}$Mg + $^{248}$Cm are larger than those for the reaction $^{36}$S + $^{238}$U. The calculated results show the same behavior. One can see that the fusion probability decreases with the decreasing mass asymmetry for the reactions $^{26}$Mg + $^{248}$Cm, $^{36}$S + $^{238}$U, and $^{48}$Ca + $^{226}$Ra. The same trend can be seen in Ref. \cite{Liu02,Hong01}. Table \ref{tab:table1} shows the quasifission barriers $B_{\textrm{qf}}$, inner fusion barriers $B_{\textrm{fus}}$, potential energy at the bottom of pocket $V(R_{m})$, and Coulomb barriers $V_{B}$ for different reactions. For the reactions $^{26}$Mg + $^{248}$Cm, $^{36}$S + $^{238}$U, and $^{48}$Ca + $^{226}$Ra, it is noticed that the $B_{\textrm{fus}}$ increases and $B_{\textrm{qf}}$ decreases with the decrease of mass asymmetry. This is the main reason for the behavior of $P_{\textrm{CN}}$ observed above.

\begin{center}
\includegraphics[width=6cm]{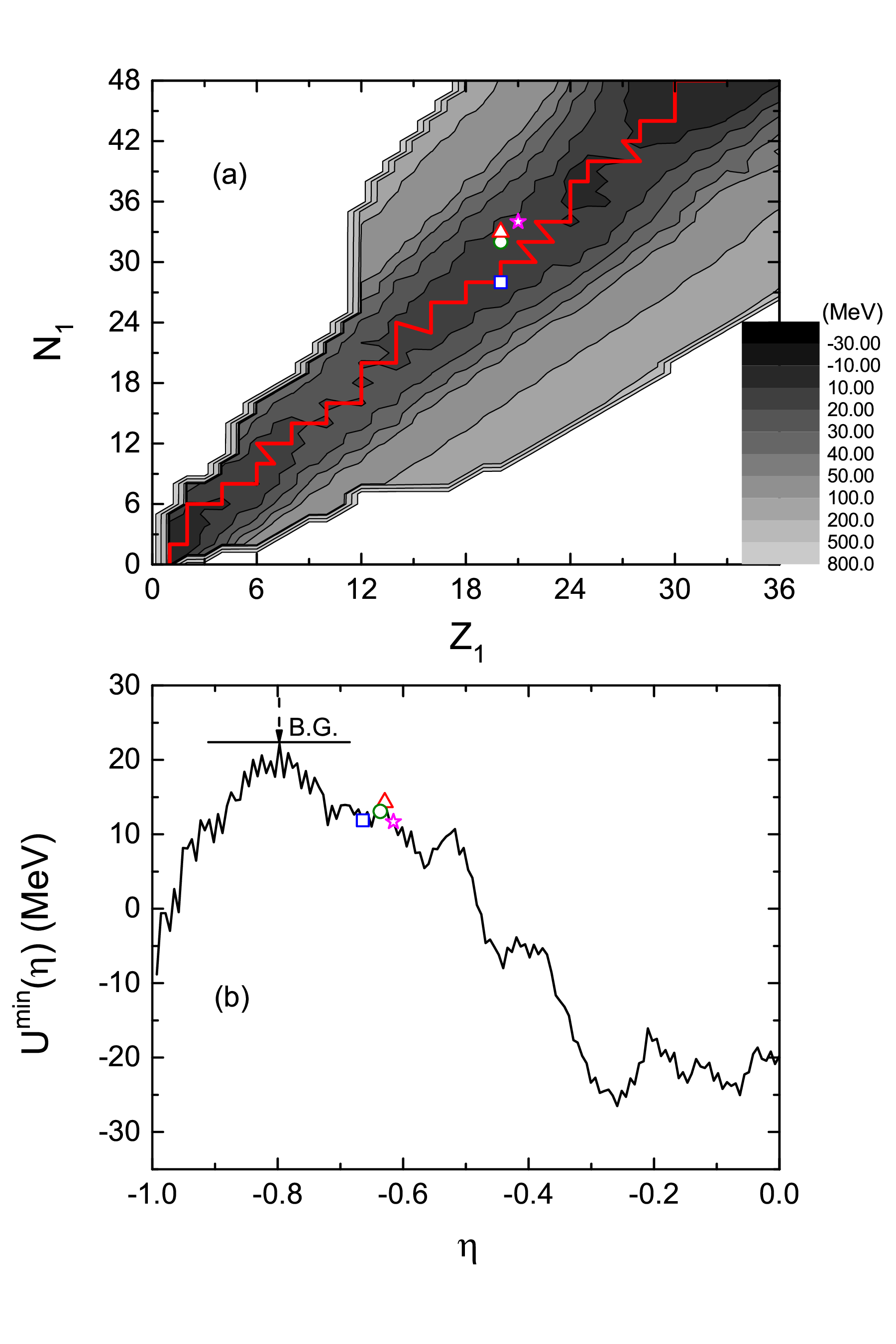}
\figcaption{\label{driver286} (Color online) The PES as functions of $Z_{1}$ and $N_{1}$ of fragments 1 for the DNS configurations leading to the compound nucleus $^{286}$Cn$^{*}$. The injection points for the reactions $^{53}$Ca + $^{233}$U, $^{52}$Ca + $^{234}$U, $^{48}$Ca + $^{238}$U, and $^{55}$Sc + $^{231}$Pa are denoted by triangle, circle, square, and star, respectively. }
\end{center}

\subsection{Fusion probabilities in formation of the compound nucleus $^{286}$Cn$^{*}$}

To clarify the entrance channel effects on $P_{\textrm{CN}}$, we also apply the model to calculate the fusion probabilities in the reactions $^{48}$Ca + $^{238}$U, $^{52}$Ca + $^{234}$U, and $^{53}$Ca + $^{233}$U, and $^{55}$Sc + $^{231}$Pa. These reactions are all in formation of the same compound nucleus $^{286}$Cn$^{*}$.

Figure \ref{driver286}(a) shows the PES as functions of proton and neutron number of fragments 1 for DNS configurations leading to the compound nucleus $^{286}$Cn$^{*}$. The injection points for the reactions $^{53}$Ca + $^{233}$U, $^{52}$Ca + $^{234}$U, $^{48}$Ca + $^{238}$U, and $^{55}$Sc + $^{231}$Pa are denoted by triangle, circle, square, and star, respectively. It can be seen that the injection points for the reactions $^{53}$Ca + $^{233}$U and $^{55}$Sc + $^{231}$Pa are located at quite high place. The reason is that the nuclei $^{53}$Ca and $^{55}$Sc are radioactive. The minimum trajectory in PES ($U^{\textrm{min}}$) can be presented as a function of the mass asymmetry $\eta$, as shown in Fig. \ref{driver286}(b).

\begin{center}
\includegraphics[width=6cm]{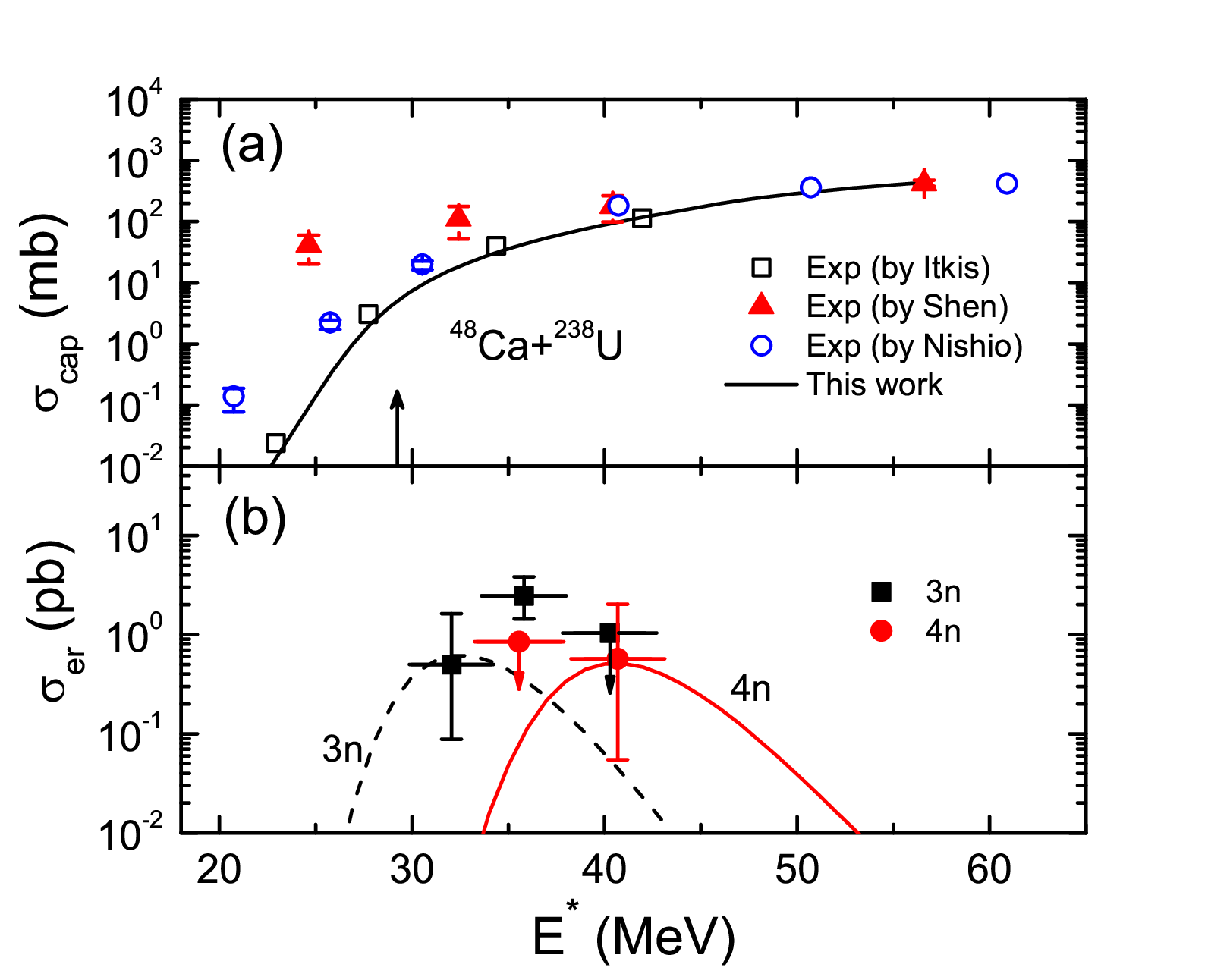}
\figcaption{\label{238u} (Color online) (a) Capture cross sections for the reaction $^{48}$Ca + $^{238}$U. The experimental data are denoted by squares \cite{Itkis02}, triangles \cite{Shen02}, and circles \cite{Nishio03}. The solid line denotes the calculated result in this work. The vertical arrow shows the position of Coulomb barrier. (b) ER cross sections for the reaction $^{48}$Ca + $^{238}$U. The dashed line and solid line denote the calculated ER cross sections of the 3n and 4n channels, respectively. The experimental data \cite{Yu04} of the 3n and 4n channels are denoted by solid squares and circles, respectively.}
\end{center}

Figure \ref{238u}(a) shows the comparison of capture cross section between results in this work and the experimental data \cite{Itkis02,Shen02,Nishio03} for the $^{48}$Ca + $^{238}$U fusion reaction. The calculated result is in good agreement with the experimental data from Ref. \cite{Itkis02} in whole incident energy region. It can be seen that calculated result also shows good agreement with all experimental data measured by Nishio \emph{et al} \cite{Nishio03} around and above the Coulomb barrier. Figure. \ref{238u}(b) shows comparison of ER cross sections between experimental data and calculated results. Within the error bars, the experimental data \cite{Yu04} are reproduced rather well, especially for the 4n channel.

\begin{center}
\includegraphics[width=6cm]{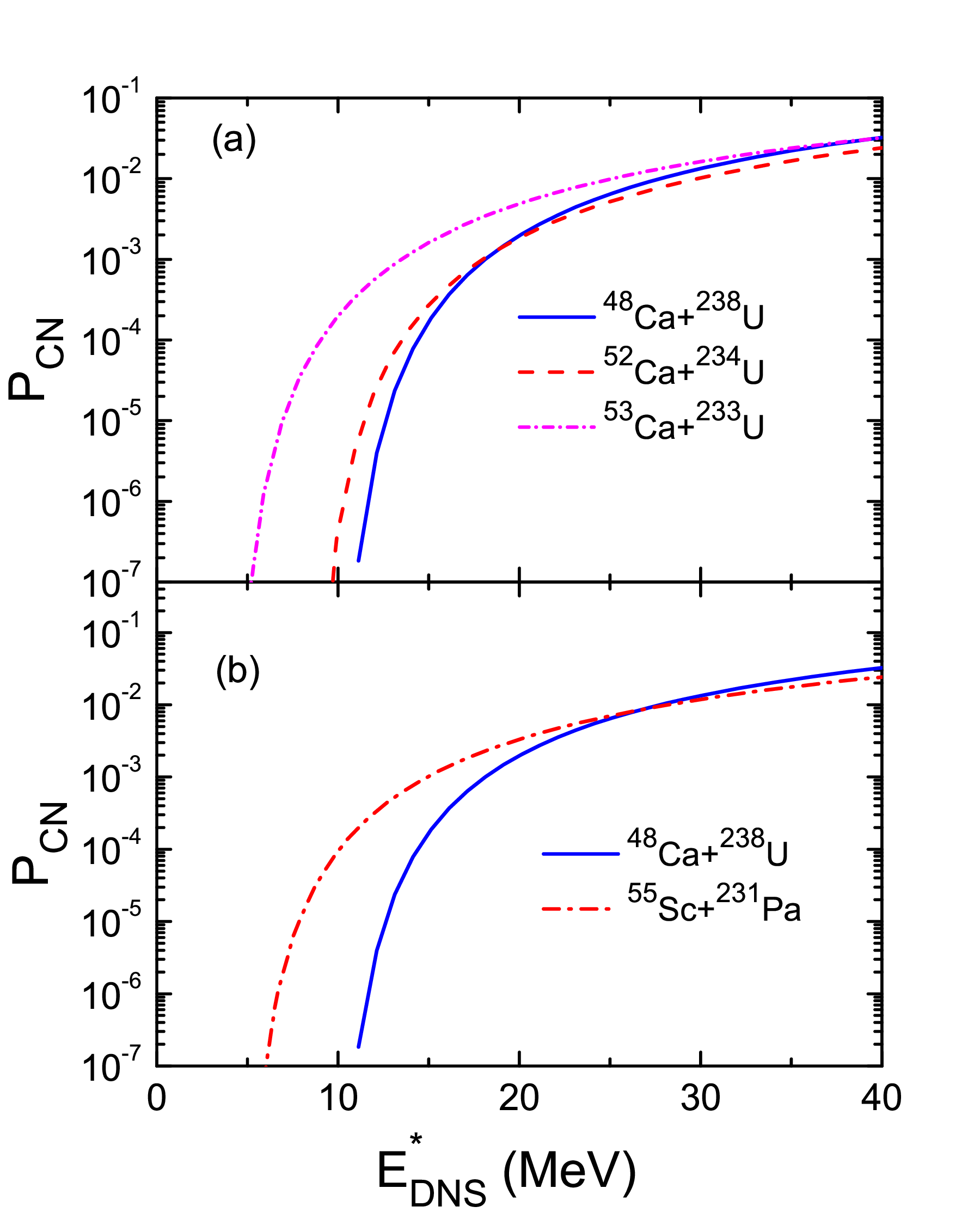}
\figcaption{\label{same-p}(Color online) (a) Calculated fusion probabilities as a function of $E^{*}_{\textrm{DNS}}$ ($J=0$) for the reactions $^{48}$Ca + $^{238}$U, $^{52}$Ca + $^{234}$U, and $^{53}$Ca + $^{233}$U with different mass asymmetry induced by neutron number. (b) Calculated fusion probabilities as a function of $E^{*}_{\textrm{DNS}}$ ($J=0$) for the reactions $^{48}$Ca + $^{238}$U and $^{55}$Sc + $^{231}$Pa with different mass asymmetry induced by proton number.
}
\end{center}

Figure \ref{same-p}(a) shows the fusion probabilities for the reactions $^{48}$Ca + $^{238}$U, $^{52}$Ca + $^{234}$U, and $^{53}$Ca + $^{233}$U. For these reactions, the difference of mass asymmetry is induced by neutron number. It can be seen that $P_{\textrm{CN}}$ for the reaction $^{53}$Ca + $^{233}$U is much larger than that for the reaction $^{48}$Ca + $^{238}$U at large $E_{\textrm{DNS}}^{*}$ range, although $^{53}$Ca + $^{233}$U shows smaller mass asymmetry. At high $E_{\textrm{DNS}}^{*}$ region, the fusion probability for the reaction $^{52}$Ca + $^{234}$U is lower than that for the reaction $^{48}$Ca + $^{238}$U, while the opposite behavior is shown at low $E_{\textrm{DNS}}^{*}$. In Fig. \ref{same-p}(b), We notice that the fusion probability of the reaction $^{55}$Sc + $^{231}$Pa is larger than that of the reaction $^{48}$Ca + $^{238}$U, although the reaction $^{55}$Sc + $^{231}$Pa has smaller mass asymmetry (induced by proton number). It is already known that the barriers $B_{\textrm{fus}}$ and $B_{\textrm{qf}}$ play crucial roles in complete fusion \cite{Fazio01}. Usually, in formation of the same compound nucleus, the fusion probability is larger for the reaction with smaller $B_{\textrm{fus}}$ and larger $B_{\textrm{qf}}$, which means a larger fusion probability for the reaction $^{48}$Ca + $^{238}$U than that for the reaction $^{55}$Sc + $^{231}$Pa (see Table \ref{tab:table1}). However, the opposite behavior is shown. Therefore, some other physical quantities need to be taken into account, especially for the radioactive beam induced hot fusion reactions.

As we discussed in Fig. \ref{driver286}, the injection point in PES for the reaction $^{52}$Ca + $^{234}$U, $^{53}$Ca + $^{233}$U, and $^{55}$Sc + $^{231}$Pa located at the higher place than the reaction $^{48}$Ca + $^{238}$U. To clarify the influence of potential energy of initial DNS on the fusion probability, we define the difference, $\Delta U=U^{\textrm{\textrm{inj}}}(Z_{1}, N_{1})-U^{\textrm{min}}(\eta)$, between the potential energy at the injection point $U^{\textrm{\textrm{inj}}}(Z_{1}, N_{1})$ (as shown in Fig. \ref{driver286}(a)) and the driving potential $U^{\textrm{min}}(\eta)$ with corresponding value of mass asymmetry at the minimum trajectory (as shown in Fig. \ref{driver286}(b)). The difference, $\Delta V(R_{\textrm{m}})=V^{\textrm{inj}}(R_{\textrm{m}})-V^{\textrm{min}}(R_{\textrm{m}})$, is also defined. Here, the superscripts ``inj" and ``min" denote the initial combinations at the injection points and configurations with the same mass asymmetry at the minimum trajectory, respectively. The difference of $Q_{\textrm{gg}}$ value, in formation of DNS configuration ($Z_{1}$, $N_{1}$; $Z_{2}$, $N_{2}$), between initial DNS configurations at injection points and corresponding configurations with same mass asymmetry at minimum trajectory can be written as $\Delta Q_{\textrm{gg}}=B_{\textrm{DNS}}^{\textrm{inj}}-B_{\textrm{DNS}}^{\textrm{min}}$. Therefore, $\Delta U=\Delta Q_{\textrm{gg}}+\Delta V(R_{\textrm{m}})$.

\begin{center}
\includegraphics[width=6cm]{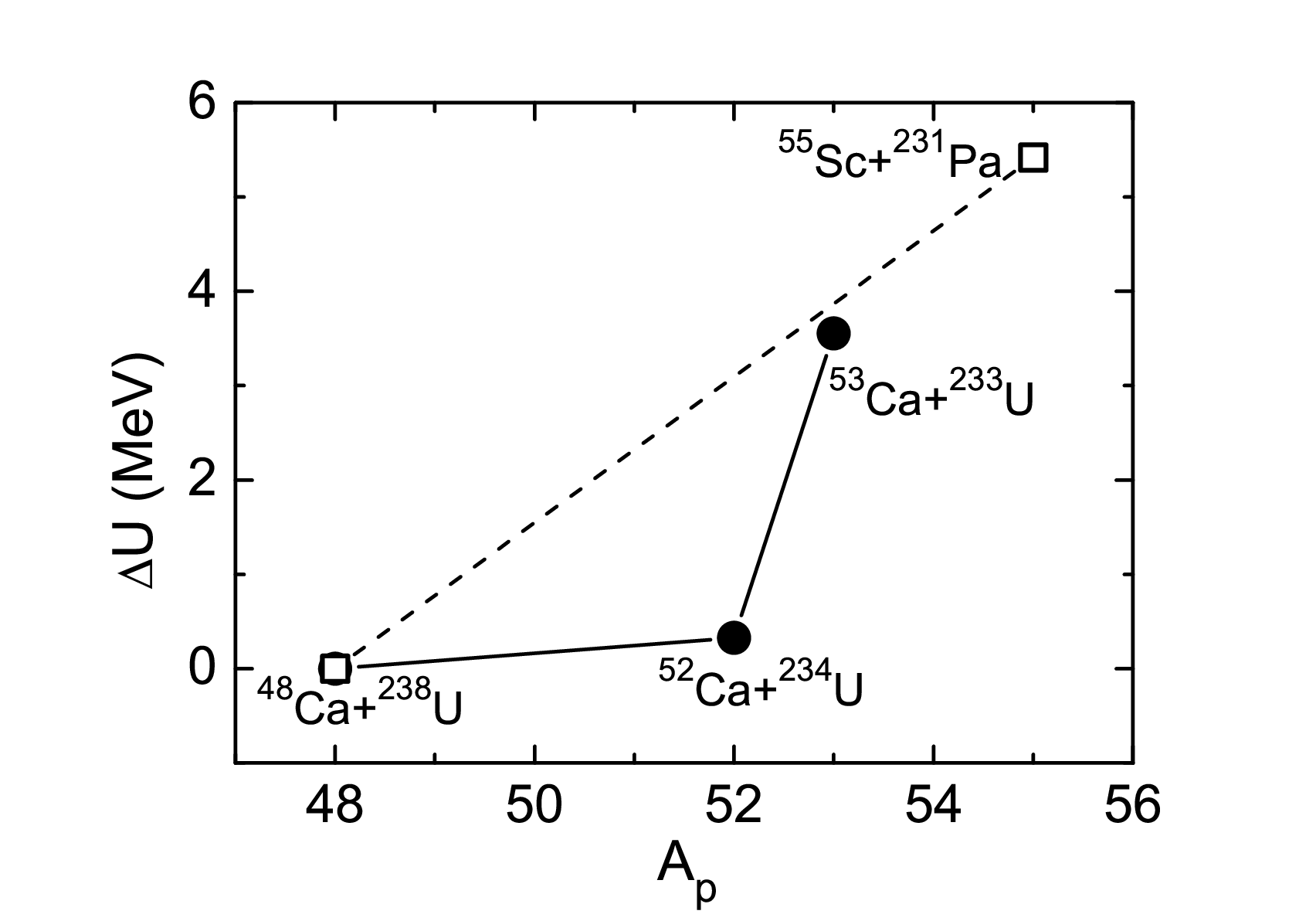}
\figcaption{\label{driv2} $\Delta U$ for the combinations leading to the same compound nucleus $^{286}$Cn$^{*}$. The circles and squares denote the reactions with different mass asymmetry induced by neutron number and proton number, respectively. The lines are used to guide the eye. }
\end{center}

We show $\Delta U$ for the reactions $^{48}$Ca + $^{238}$U, $^{52}$Ca + $^{234}$U, $^{53}$Ca + $^{233}$U, and $^{55}$Sc + $^{231}$Pa in Fig. \ref{driv2}. It can be seen that $\Delta U$ for the reactions $^{53}$Ca + $^{233}$U and $^{55}$Sc + $^{231}$Pa is much larger than that for the reaction $^{48}$Ca + $^{238}$U. Apart from the local excitation energy of forming a DNS configuration ($Z_{1}$, $N_{1}$; $Z_{2}$, $N_{2}$) in the corresponding configuration with same mass asymmetry located at the minimum trajectory, some extra excitation energy ($=\Delta U$) will be released in the reactions $^{52}$Ca + $^{234}$U, $^{53}$Ca + $^{233}$U, and $^{55}$Sc + $^{231}$Pa, as shown in Eq. (\ref{local}). The extra excitation energies in the reactions $^{53}$Ca + $^{233}$U and $^{55}$Sc + $^{231}$Pa are much larger than that in the other reactions. Therefore, with the same $E_{\textrm{DNS}}^{*}$, in formation of the same DNS configuration ($Z_{1}$, $N_{1}$; $Z_{2}$, $N_{2}$) the local excitation energy $E_{1}$ for the reaction $^{53}$Ca + $^{233}$U and $^{55}$Sc + $^{231}$Pa are larger than others. As we known, the fusion probability is sensitive to the value of the excitation energy. The $P_{\textrm{CN}}$ usually increases with the increasing $E_{1}$ \cite{Feng01}.  Therefore, the enhancement of fusion probability for the reactions $^{53}$Ca + $^{233}$U and $^{55}$Sc + $^{231}$Pa are mainly due to large $\Delta U$.

\section{Conclusions}

In this article, we have investigated the entrance channel effects of fusion probability within the DNS model.
The reactions $^{26}$Mg + $^{248}$Cm, $^{36}$S + $^{238}$U, $^{48}$Ca + $^{226}$Ra, $^{53}$Ca + $^{233}$U, $^{55}$Sc + $^{231}$Pa, $^{48}$Ca + $^{238}$U, and $^{52}$Ca + $^{234}$U leading to the compound nuclei $^{274}$Hs$^{*}$ and $^{286}$Cn$^{*}$ are studied.
It has been claimed that the fusion probability is enhanced for a mass asymmetric reaction \cite{Adamian04}.
However, our calculations show that the fusion probabilities as a function of excitation energy of DNS for the reactions $^{53}$Ca + $^{233}$U and $^{55}$Sc + $^{231}$Pa are larger than that for the $^{48}$Ca + $^{238}$U reaction, although the mass asymmetry of $^{48}$Ca + $^{238}$U is larger.
The enhancement of fusion probability is mainly due to large value of potential energies of initial DNS.
For the future work, the capture cross sections and survival probability also need to be consider in finding favorable radioactive induced hot fusion reactions to synthesize SHN.
\end{multicols}

\vspace{-1mm}
\centerline{\rule{80mm}{0.1pt}}
\vspace{2mm}
\begin{multicols}{2}

\end{multicols}

\clearpage
\end{CJK*}
\end{document}